\documentclass[prl,showpacs,showkeys,preprintnumbers,amsmath,amssymb,superscriptaddress,twocolumn]{revtex4}

\usepackage{graphicx}
\usepackage{amsfonts}

\def\l{\ell}
\def\lx{\l_x}
\def\ly{\l_y}
\def\lz{\l_z}

\def\vup{v_{\textrm{shear}}}

\def\e{\cdot 10^}

\def\eps{\varepsilon}

\begin{document}

\title{ 
Flow, Ordering and Jamming of Sheared Granular Suspensions\\
}

\author{Denis~S.~Grebenkov}
 \email{denis.grebenkov@polytechnique.edu}

\affiliation{Dip.to di Scienze Fisiche,
 Universit\'a di Napoli ``Federico II'' and INFN, Naples,
ITALY}

\affiliation{
LPMC,
C.N.R.S. -- Ecole Polytechnique, F-91128
Palaiseau, FRANCE }

\author{Massimo Pica Ciamarra}
\affiliation{CNISM \& Dip.to of Information Engineering,
 Seconda Universit\'a di Napoli, Aversa, ITALY}
\affiliation{Dip.to di Scienze Fisiche,
 Universit\'a di Napoli ``Federico II''  and INFN, Naples,
ITALY}

\author{Mario Nicodemi}
\affiliation{Dip.to di Scienze Fisiche,
 Universit\'a di Napoli ``Federico II''  and INFN, Naples,
ITALY}

\affiliation{Complexity Science \& Department of Physics,
University of Warwick, UK}

\author{Antonio Coniglio}

\affiliation{Dip.to di Scienze Fisiche,
 Universit\'a di Napoli ``Federico II''  and INFN, Naples,
ITALY}

\date{Received: \today / Revised version: }

\begin{abstract}
We study the rheological properties of a granular suspension subject
to constant shear stress by constant volume molecular dynamics
simulations.  We derive the system `flow diagram' in the volume
fraction/stress plane $(\phi,F)$: at low $\phi$ the flow is
disordered, with the viscosity obeying a Bagnold-like scaling only at
small $F$ and diverging as the jamming point is approached; if the
shear stress is strong enough, at higher $\phi$ an ordered flow regime
is found, the order/disorder transition being marked by a sharp drop
of the viscosity.  A broad jamming region is also observed where, in
analogy with the glassy region of thermal systems, slow dynamics
followed by kinetic arrest occurs when the ordering transition is
prevented.
\end{abstract}




\maketitle 

Under the effects of external drives granular media exhibit a variety
of complex dynamical behaviors.  Compaction under shaking is a well
studied phenomenon, characterized by slow relaxation followed by a
jamming transition at high volume fraction
\cite{Knight,NCH,Richard} with deep analogies to thermal glasses
\cite{NCH,Liu98,Coniglio04,Richard}. In addition shear-induced
transitions from flowing to jammed states \cite{NagelNat05} or
to ordered flowing states are observed \cite{Tsai03,DanielsPRL}.
In such a complex panorama, the rheology of the different
phases of a system under shear is far from being clarified,
and even a clear `flow diagram' locating the regions with
ordered flow, disordered flow, and jamming, is missing.

Within such a perspective, we investigate by molecular dynamics (MD)
simulations the rheology of a dense granular suspension subject to a
constant shear stress in a box of constant volume. Alike previous
experiments on non-Brownian particles (see \cite{Bertrand,Pine05b} and
references therein), we opt for a constant shear stress rather that a
constant shear rate setup (see \cite{Tsai03} and references therein)
because jamming may be precluded in the latter case.  
We derive the system `flow diagram' as obtained by shearing for a long time 
a disordered assembly of grains. It includes an order/disorder transition line, signaled 
by a sharp drop of the effective viscosity $\eta$, as well as a jamming
region. Similarly to supercooled liquids, a crossover from a
transient disordered flow to a stationary ordered one is also found,
jamming occurring when ordering is kinetically impeded.  In the regime
of disordered flow, the system viscosity, $\eta$, is found
to grow with the applied shear stress according to Bagnold scaling,
but it departs from it when the shear stress increases.  As the
packing fraction grows towards a critical value, $\eta$ is found
to diverge with a power law. We also show that, similarly to glassy
thermal systems, the long time state of the system in the jamming
region is dependent on the dynamical preparation protocol.

We consider monodisperse spherical beads of diameter $d$ and mass $m$,
initially randomly located between the upper and lower plates of a box
of a given size $\l_x \times \l_y\times \l_z$.  The lower plate is
immobile and the upper one may move along $x$ in response to a given
applied shear stress $\sigma$, i.e., under a constant force $F =
\sigma \lx \ly$ (periodic boundary conditions are used along the $x$
and $y$ axes).  Grains are studied by MD simulations of a well known
linear spring-dashpot model (L3 of Ref.~\cite{Silbert01}), including
particle rotations and static friction: normal interaction between two
beads is characterized by the elastic and viscoelastic constants $k_n
= 2\e{5}$ and $\gamma_n = 50$, given in standard units
\cite{units}; static friction is implemented by keeping track of the
elastic shear displacement throughout the lifetime of a contact.  The
tangential elastic constant is $k_t=5.7\e{4}$; the static friction
coefficient $\mu=0.1$; the restitution coefficient is $0.88$.  The box
top and bottom plates are made of spherical beads, randomly `glued' at
height $z = \lz - 1/2 + \eps$ and at $z = 1/2 + \eps$, respectively,
where $\eps$ is randomly chosen for each bead in the set $[-1/4,1/4]$.
The area fraction of grains on the plates is high (large non-physical
overlaps are allowed between the constituent particles) and their
mass is $300m$.  We explored a range of box sizes, and below we
illustrate the results by considering the case with $\lx = \ly = 16$,
and $\lz = 8$ where finite size effects are absent.  The small value of $\lx$ and $\ly$,
taken here to allow simulations on time scales long enough, is
partially compensated by our periodic boundary conditions 
($l_z$ has a value typical for experiments
\cite{Tsai03,DanielsPRL,Pine05b}).  To avoid
gravity-induced compaction, gravitational acceleration of grains is
neglected as in density-matching liquid experiments~\cite{Pine05b}.

\begin{figure}
\begin{center}
\includegraphics*[scale=0.33]{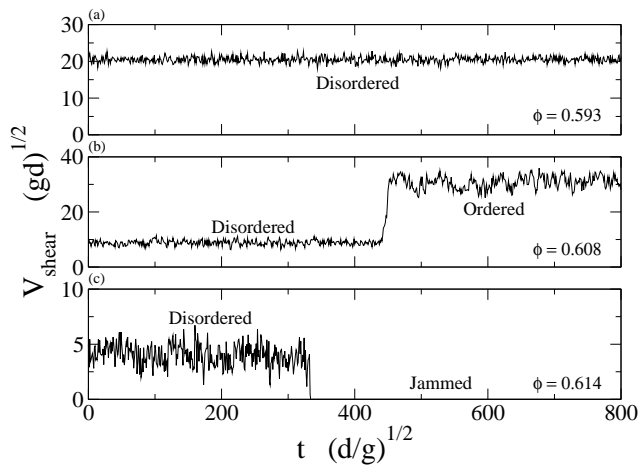}
\end{center}
\caption{
Velocity of the system upper plate, $\vup$, as a function of time,
$t$, for the shown values of the volume fraction, $\phi$, sheared
with a constant force $F = 4\cdot 10^4$.  The system starts from a
disordered initial pack.  {\bf Panel (a)} When $\phi$ is small enough
the system flows in a disordered configuration (see text and
Fig.~\ref{fig:snapshot} left panel). {\bf Panel (b)} When $\phi$ is
increased the disordered flow is only transient, as the system has an
abrupt transition to an ordered flow with a reduced viscosity (see
text and Fig.~\ref{fig:snapshot} right panel). {\bf Panel (c)} At even
higher $\phi$ the transient disordered flow has a transition to a
jammed configuration where $\vup=0$.  }
\label{fig:velocity}
\end{figure}

For each value of the grain volume fraction $\phi$ and of the applied force $F$ 
we have performed $10$ simulations using 
different initial disordered states, always finding the same typical velocities as well as 
the same long time state (we have performed simulations up to $t = 8000$):
the system rheology is therefore characterized by the value of $F$
and on $\phi$ (or equivalently on grain
number $N$ \footnote{Due to the box rough boundaries (and finite size
of the investigated system) the volume fraction definition has a
degree of ambiguity.  We use $\phi = N v_g/(\lx \ly \lz)$, where $v_g$
is the volume of a bead.  Other definitions, as for instance $\phi' =
N v_g/(\lx \ly \lz - V_{b})$, where $V_{b}$ is the volume of the beads
forming the rough boundaries, may differ up to $10\%$.}).  
In the $\phi$ and $F$ range of values we explored, which is close to the hard-particle limit
as the maximum deformation of a particle is $\delta d/d < 10^{-2}$,
three qualitatively different regimes are usually observed in the system dynamics,
corresponding to different behaviors of the measured upper plate
velocity, $\vup(\phi,F)$.  They are summarized in
Fig.~\ref{fig:velocity} which shows $\vup$ as a function of time, $t$,
at increasing values of the volume fraction, $\phi = 0.593, 0.608$ and
$ 0.614$, for $F = 4\cdot 10^4$. 

\begin{figure}
\begin{center}
\includegraphics*[scale = 1.1]{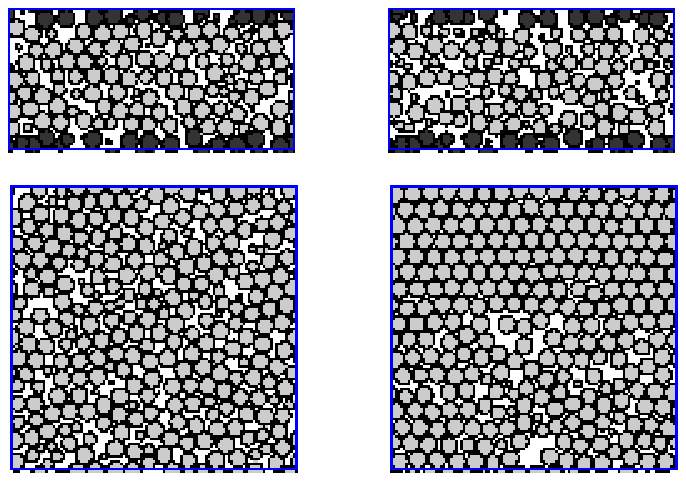}
\begin{picture}(0,0)(0,0)
\put(-230,145){\large z}
\put(-109,145){\large z}
\put(-140,100){\large x}
\put(-19,100){\large x}
\put(-140,-5){\large x}
\put(-19,-5){\large x}
\put(-230,90){\large y}
\put(-109,90){\large y}
\put(-116,0){\line(0,1){152}}
\end{picture}
\end{center}
\caption{Vertical ($ZX$, top) and horizontal ($YX$, bottom)
snapshots of sections of the system for the run of
Fig.~\ref{fig:velocity}b.  The left and the right panels show
sections taken at t = 400, when the system flow is disordered, and
at $t = 800$, when the flow is ordered. The ordered state is
characterized by the formation of crystal-like layers in the $XY$
plane, shifted along $z$.  Dark beads form the boundaries (the shown
circles have various sizes as they are different cuts of our
monodisperse spheres). } \label{fig:snapshot}
\end{figure}

When $\phi$ is small enough
(Fig.~\ref{fig:velocity}a), the system flows in a stationary
disordered state (as the one depicted in the left panel of
Fig.\ref{fig:snapshot}) from the initial random configuration and
$\vup$ fluctuates around a constant value depending on $\phi$ and $F$.
The degree of ordering of the system is
usually quantified by the amplitude of the peak of the structure
factor $S(k)$ for, say, $k = (0,0,2\pi/d)$; in the disordered region,
$S(k)$ has no peaks at all and takes values typical to disordered
arrangements, as shown in Fig.\ref{fig:diagram}b. 

At higher $\phi$ values, when $F$ is strong enough
(Fig.~\ref{fig:velocity}b), the system is trapped for a certain time
in a transient state where $\vup$ keeps a constant value up to a
moment when it suddenly jumps to a higher stationary plateau.
Correspondingly, the system exits the disordered flowing state 
(left panel of Fig.\ref{fig:snapshot}) and develops a
layered and partially ordered structure (shown
in the right panel of Fig.\ref{fig:snapshot}).  

The high $\phi$ scenario is drastically changed if the driving force
$F$ is not strong enough (see Fig.~\ref{fig:velocity}c): after a
transient flow the system jams in a state as disordered as the initial
one, and the upper plate velocity becomes zero.  The above
observations highlight that disordered states can be either
stationary, or transient.  The latter states can undergo
transitions towards ordered flowing states, signalled by a marked
increase of the shear velocity, or alternatively freeze towards jammed
packs.  

\begin{figure}
\begin{center}
\includegraphics*[scale=0.33]{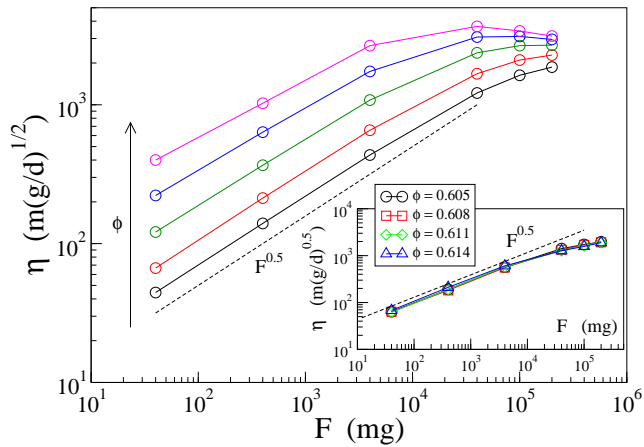}
\end{center}
\caption{The effective viscosity, $\eta=F/\vup(F)$, as a function of
the applied external force $F$ for different values of the sample
volume fraction ($\phi = 0.573-0.619$), recorded in the disordered
flow regime.  At small $F$ the viscosity increases, at a good
approximation, as a power law in $F^{\alpha}$, with $\alpha\simeq
0.5$, which implies Bagnold scaling $\sigma \propto
\dot\gamma^{\beta}$, with $\beta\simeq 2$.  {\bf Inset:} A similar
power law scaling is found for $\eta(F)$ is the ordered flow regime,
but almost no dependence with $\phi$ is observed. }
\label{fig:viscosity_F}
\end{figure}

As the velocity profiles appear to be approximately linear in
the considered cases, the system resistance to flow can be quantified
via an effective viscosity $\eta = F/\vup(\phi,F)$.  In
Fig.~\ref{fig:diagram}c we depict the `flow diagram' of the
suspension in the $(\phi,F)$-plane.
We first consider the flow in the disordered region.
Figure~\ref{fig:viscosity_F} shows the dependence of the effective
viscosity on the applied shear force $F$, for several values of the
volume fraction of the system, $\phi$.  At low $\phi$ and small
applied shear stress, the viscosity increases to a good approximation
with a power law, $\eta \propto F^\alpha$, where the exponent is
$\alpha\simeq 0.5$.
Accordingly, the relation between shear stress ($\sigma \propto F$)
and shear rate ($\dot \gamma$ $\propto \vup$) is $\sigma \propto
\dot\gamma^{\beta}$, with $\beta = 1/(1-\alpha) \simeq 2$, in good
agreement with the prediction of Bagnold scaling ($\beta  = 2$).

\begin{figure}
\begin{center}
\includegraphics*[scale=0.33]{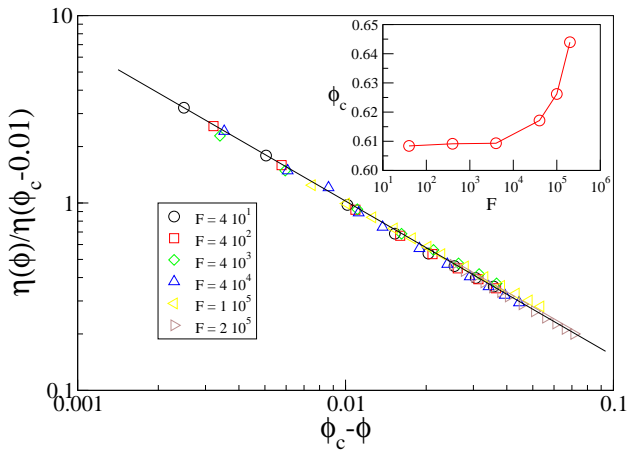}
\end{center}
\caption{ At a given force $F$, the effective viscosity measured in
the disordered flow regime diverges as a power law as the volume
fraction of the sample approaches a jamming critical threshold
$\phi_c(F)$. Interestingly, data collected at several $F$ values
collapse on a master power law, suggesting that its exponent is $F$
independent. Here, we have scaled the data in such a way that 
$\eta(\phi_c - 0.01) = 1$. 
{\bf Inset:} the fitted value to the critical volume
fraction, $\phi_c$, increases with the applied external shear
force.} \label{fig:viscosity}
\end{figure}

Within the disordered region, for a given force $F$, the dynamics of
the system strongly slows down as $\phi$ increases.  This is apparent
from Fig.~\ref{fig:viscosity} where we show that, at constant $F$, the
effective viscosity diverges with a power law $\eta(\phi) \propto
(\phi_c(F)-\phi)^{-b}$ as the packing fraction increases.  Such a
relation recorded in the disordered regime anticipates the presence of
a jamming critical volume fraction $\phi_c(F)$ (see
Fig.~\ref{fig:viscosity}, inset).
Interestingly, the fit of $\phi_c(F)$ from the small $\phi$-value region
always overestimates the volume fraction where we indeed observe the
system to jam (see Fig.~\ref{fig:diagram}), in analogy to glass
forming materials \cite{Stillinger}.
Fig.~\ref{fig:viscosity} also shows that data recorded at different
$F$ collapse one onto the other when plotted as a function of
$\phi_c(F)-\phi$, suggesting that the exponent $b\simeq 0.8$ is
independent of $F$.  This kind of power law dependence is frequently
found in colloidal systems under shear~\cite{krieger}, with the
exponent $b$ varying between $1$ and $2$.

\begin{figure}
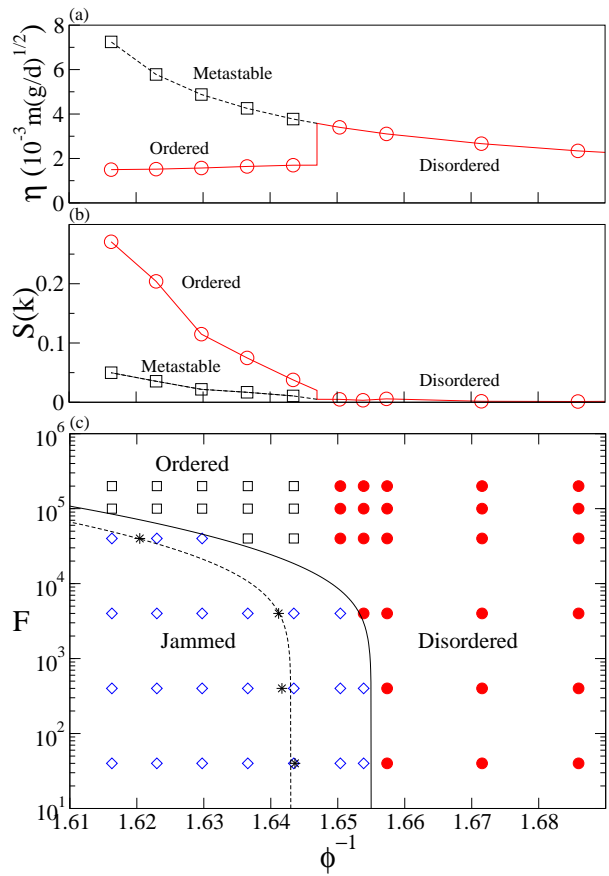

\begin{center}
\includegraphics*[scale=0.332]{jumps2.eps}\\
\includegraphics*[scale=0.33]{dia.eps}
\end{center}
\caption{ {\bf Panel (a)} and {\bf (b)} show the dependence of the
effective viscosity $\eta$ and of the structure factor $S(k)$ (for $k
=(0,0,2\pi/d)$) on the inverse volume fraction, $\phi^{-1}$, for $F =
10^5$. In the low $\phi$ regime the system flow is disordered as
$S(k)$ has typically small values. By increasing $\phi$, there is an
initial transient disordered flow ($S(k)$ small) with growing $\eta$,
followed by a sharp transition to an ordered flow (with a much higher
$S(k)$, see text) with a sharp drop in the viscosity (see
Fig.~\ref{fig:velocity}b). {\bf Panel (c):} The system `flow diagram'
showing the long time flow regime as a function of the inverse volume
fraction $\phi^{-1}$ and of the applied force $F$: circles mark the
region of disordered flow, open squares mark ordered flow, and diamonds
the jammed region.  Stars indicate the inverse critical volume
fraction $\phi_c^{-1}$ defined from the fit in
Fig.~\ref{fig:viscosity}.
}
\label{fig:diagram}
\end{figure}

We now consider the regime observed at higher $\phi$ where, under
shearing, the viscosity exhibits a sharp transition from a
disordered to a faster ordered flow
as in Fig~\ref{fig:velocity}b.
The change to the faster flow, with the corresponding reduction
of the effective viscosity, is associated to the formation of
partially ordered layers parallel to the shearing plate
(Fig.~\ref{fig:snapshot}), as previously observed in experiments on
granular and colloidal suspensions~\cite{Tsai03,Tsai04,ordering}. 

The connection between ordering and viscosity reduction is apparent
in Fig.~\ref{fig:diagram}a,b, where we show the
dependence of both $\eta$ and $S(k)$ on the inverse volume
fraction of the system (squares indicate values recorded in the
transient regime, circles the asymptotic ones).
In the low $\phi$ region, before the ordering transition, $S(k)$
has small values typical to disordered arrangements.
At higher $\phi$, the system can flow disordered, with increasing
viscosity and $S(k)$ having still comparable small values.
Eventually, when a sharp drop in the viscosity is observed, it is
accompanied by a clear increase of the structure factor $S(k)$,
pointing out the presence of order in the system, as seen in the right
panel of Fig.~\ref{fig:snapshot}. In the ordered flow region, $\eta$
is almost insensitive to $\phi$ in the range here explored (see
Fig.\ref{fig:diagram}a) and has an approximate Bagnold-like scaling in
$F$, $\eta\propto F^{\alpha}$, with $\alpha\simeq 0.5$ (see inset in
Fig.\ref{fig:viscosity_F}).In the high $\phi$ regime, if the applied shear stress is not strong
enough, the initial transient disordered flow does not generally
result in an ordered stationary flow, since a full dynamical arrest
(see Fig.~\ref{fig:diagram}c) with $\vup=0$ is found.
Jamming occurs when the system disordered configurations are
kinetically trapped in states where further shear (at the given value
of $F$) becomes impossible, the underlying mechanism being still
unclear~\cite{Cates,Ball}.  Such a trapping is broken when $F$ is high
enough and ordered flow appears.

The flow diagram of Fig.~\ref{fig:diagram}c summarizes the 
properties of the long time states reached by
the system when the initial condition is disordered,
and suggests an analogy between the $(\phi,F)$ flow diagram 
and the $(\phi,T)$ phase diagram of usual thermal systems, the disordered, ordered and jammed
states corresponding respectively to the liquid, crystalline and
glassy phases. To reinforce this analogy we have investigated 
the behavior of a system initially prepared in an ordered configuration.
Interestingly, we have found that in the region of low $F$ and high $\phi$, 
where jamming was previously found,
the ordered flow is not arrested and keeps going with a 
finite value of $\eta$ (although the system might jam at times longer than those we can investigate).
In the other regions of the flow diagram, on the contrary, the long time state of the system is
that shown in Fig.~\ref{fig:diagram}c regardless of the initial conditions.
This dependence on the initial conditions is analogous to that observed in thermal systems,
where the cristalline phase is stable in the region where the glassy phase is found.
\\
\\
\noindent
This work has been supported by EU Network MRTN-CT-2003-504712.


\begin{thebibliography}{30}
\bibitem{units}
Distances, times, velocities, forces, elastic constants, and stresses are, respectively, measured in
units of $d$, $t_0 = \sqrt{d/g}$, $v_0 = \sqrt{gd}$, $F_0 = mg$, $k_n= mg/d$,
and $\sigma_0 = mg/d^2$, where $g$ is gravity acceleration.
\end{thebibliography}
\end{document}